\documentclass[12pt,document,nofootinbib,superscriptaddress,onecolumn,preprintnumbers,balancelastpage]{revtex4}
\pdfoutput=1
\usepackage{latexsym}
\usepackage{amssymb}
\usepackage{epsfig,amsmath,graphics}
\usepackage{listings}
\usepackage{color}
\usepackage{dcolumn}
\usepackage{slashed}
\usepackage{cancel}
\usepackage{comment,latexsym} 
\usepackage{bm}
\usepackage{verbatim}
\usepackage{tabularx}
\usepackage{dcolumn}
\definecolor{Mahogany}{rgb}{0.62,0.24,0.15}
\definecolor{colorLink}{rgb}{0.7,0,0}
\definecolor{colorCite}{rgb}{0,.7,0}
\definecolor{colorURL}{rgb}{0,0,0.7}
\usepackage[colorlinks=true,linktocpage=true,linkcolor=black,citecolor=colorCite,urlcolor=colorURL]{hyperref}
\usepackage{setspace}
\usepackage{xspace}
\usepackage{multirow}
\usepackage{titlesec}

\def\be{\begin{equation}}
\def\ee{\end{equation}}
\newcommand{\beq}{\begin{equation}}
\newcommand{\eeq}{\end{equation}}
\newcommand{\Eref}[1]{Eq.~(\ref{#1})}
\newcommand{\Erefs}[2]{Eqs.~(\ref{#1}) and~(\ref{#2})}
\newcommand{\Tref}[1]{Table~\ref{#1}}
\newcommand{\Sref}[1]{Sec.~\ref{#1}}
\newcommand{\Fref}[1]{Fig.~\ref{#1}}

\newcommand{\lsim}{\!\mathrel{\hbox{\rlap{\lower.55ex \hbox{$\sim$}} \kern-.34em \raise.4ex \hbox{$<$}}}}
\newcommand{\gsim}{\!\mathrel{\hbox{\rlap{\lower.55ex \hbox{$\sim$}} \kern-.34em \raise.4ex \hbox{$>$}}}}
\newcommand{\vev}[1]{ \left\langle {#1} \right\rangle }
\newcommand{\GeV}{{\text{ GeV}}}
\newcommand{\TeV}{{\text{ TeV}}}

\newcommand{\eg}{\emph{e.g.}}
\newcommand{\abs}[1]{\left|#1\right|}

\expandafter\def\expandafter\normalsize\expandafter{%
    \normalsize
    \setlength\abovedisplayskip{8pt}
    \setlength\belowdisplayskip{8pt}
    \setlength\abovedisplayshortskip{8pt}
    \setlength\belowdisplayshortskip{8pt}
}

\titleformat{\section}{\center\normalfont\fontsize{14}{15}\bfseries}{\thesection.}{1em}{}
\titleformat{\subsubsection}{\center\normalfont\fontsize{12}{15}}{\thesubsubsection.}{1em}{}

\begin{document}

\begin{flushright}
\text{\footnotesize MCTP-15-02}\\
\vskip -6 pt
\text{\footnotesize FERMILAB-PUB-15-003-T}\\
\end{flushright}
\vskip 10 pt

\title{\Large Natural Supersymmetry without Light Higgsinos} 

\author{\large Timothy Cohen}
\affiliation{
Department of Physics, Princeton University, Princeton, NJ, 08544 \\
School of Natural Sciences, Institute for Advanced Study, Princeton, NJ, 08540\\
Institute of Theoretical Science, University of Oregon, Eugene, OR, 97403\\
Theory Group, SLAC National Accelerator Laboratory, Menlo Park, CA, 94025}

\author{\large John Kearney}
\affiliation{
Theoretical Physics Department \\
\vskip -8 pt
Fermi National Accelerator Laboratory, Batavia, IL 60510\\
Michigan Center for Theoretical Physics\\
\vskip -8 pt
University of Michigan, Ann Arbor, MI 48109}

\author{\large Markus A.~Luty}
\affiliation{
Physics Department, University of California Davis, Davis, CA 95616}

\begin{abstract}
\vskip 1 pt
\begin{center}
{\bf Abstract}
\end{center}
\vskip -30 pt
$\quad$
\begin{spacing}{1.1}\noindent
We present a mechanism that allows a large Higgsino mass without large fine-tuning.  The Higgs is a pseudo Nambu-Goldstone boson (PNGB) of the global symmetry breaking pattern $SO(5) \to SO(4)$.  Because of the PNGB nature of the light Higgs, the $SO(5)$ invariant Higgsino mass does not directly contribute to the Higgs mass.  Large couplings in the Higgs sector that spontaneously breaks $SO(5)$ minimize the tuning, and are also motivated by the requirements of generating a sufficiently large Higgs quartic coupling and of maintaining a natural approximate global $SO(5)$ symmetry.  When these conditions are imposed, theories of this type predict heavy Higgsinos.  This construction differs from composite Higgs models in that no new particles are introduced to form complete $SO(5)$ multiplets involving the top quark---the stop is the only top partner.  Compatibility with Higgs coupling measurements requires cancelations among contributions to the Higgs mass squared parameter
at the $10\%$ level.  An important implication of this construction is that the compressed region of stop and sbottom searches can still be natural.
\end{spacing}
\end{abstract}


\maketitle
\newpage
\begin{spacing}{1.3}
\pagebreak
%
%
\newcommand{\Eq}[1]{Eq.~(\ref{#1})}
\newcommand{\Eqs}[1]{Eqs.~(\ref{#1})}
\newcommand{\eq}[1]{(\ref{#1})}
\newcommand{\Ref}[1]{Ref.~\cite{#1}}
\newcommand{\Refs}[1]{Refs.~\cite{#1}}
\newcommand{\la}{\lambda}

\section{Introduction}
\label{sec:Intro}
The Standard Model of particle physics is in many respects the perfect effective quantum field theory.  It is fully determined by including all possible relevant and marginal couplings compatible with the particle content, Lorentz symmetry, and gauge invariance.  The resulting theory accurately describes all interactions of elementary particles up to the highest energies probed by experiment, including the intricate structure of electroweak interactions and flavor-changing transitions;  allowing dimension-5 couplings suppressed by a large mass scale accounts for small neutrino masses and oscillations.  Moreover, the recent discovery of the 125 GeV Higgs by the ATLAS \cite{Aad:2012tfa} and CMS \cite{Chatrchyan:2012ufa} collaborations at the Large Hadron Collider (LHC) has experimentally completed the Standard Model.

However, this triumph of effective field theory is marred by the fact that the Standard Model does not give us any understanding of the size of the Higgs mass, the single relevant parameter of the model (ignoring the cosmological constant).  In particular, new physics at exponentially high energies, such as the grand unification scale $\sim 10^{16}\GeV$ or the Planck scale $\sim 10^{19}\GeV$, generically gives contributions to the Higgs mass proportional to the relevant scale, requiring fine tuning of fundamental parameters to explain the observed value.  This motivates models of physics beyond the Standard Model in which the Higgs mass parameter is calculable and naturally of order the electroweak scale.

A basic but very important point about this problem is that obtaining a light Higgs mass in the Standard Model requires only a single tuning.  This is also true in models with supersymmetry (SUSY), which are the focus of this paper.  These models have many parameters, but requiring the absence of fine-tuning constrains only one combination of them.  As such, the tuning is dominated by the superpartner masses that give the largest correction to the effective Higgs mass \cite{Dimopoulos:1995mi, Cohen:1996vb}.  In the minimal supersymmetric standard model (MSSM) we obtain (see \eg~\cite{Papucci:2011wy, Brust:2011tb})
\begin{equation}
\label{tuning}
\frac{1}{\text{tuning}} \sim 5 
\times \max\left\{
\left( \frac{\mu}{200\GeV} \right)^2,\ 
\frac{m_{\tilde{t}_1}^2 + m_{\tilde{t}_2}^2}{(600\GeV)^2},\ 
\left( \frac{m_{\tilde{g}}}{900\GeV} \right)^2 
\right\},
\end{equation}
where $\mu$ is the Higgsino mass, $m_{\tilde{t}_{1,2}}$ are the masses of the two stop mass eigenstates, and $m_{\tilde{g}}$ is the gluino mass.  For simplicity we have assumed large $\tan\beta$ and neglected $A$ terms (no stop mixing).  The lack of signals in the impressive variety of SUSY searches at the 8~TeV LHC sets lower bounds on superpartner masses, pushing the theory toward the fine-tuned regime.  The large number of possible spectra makes it impossible to draw completely general conclusions regarding naturalness, but is probably fair to say that naturalness has been experimentally probed at the $10\%$ level (see \eg~\cite{Evans:2013jna}).

Another independent tension with naturalness is that the observed physical Higgs mass $m_h$ is an additional source of tuning.  In the MSSM, the observed value $m_h \simeq 125 \GeV$ requires $m_{\tilde{t}} \sim 1~\text{TeV}$, which implies percent-level tuning, as we see in \Eq{tuning}.  \Eq{tuning} neglects $A$ terms and hence stop mixing, but including this does not alleviate the tension \cite{Hall:2011aa}.  Naturalness therefore motivates extensions of the MSSM Higgs sector, and many different possibilities have been explored in the literature \cite{Batra:2003nj,Maloney:2004rc, Azatov:2011ht, Azatov:2011ps, Heckman:2011bb, Gherghetta:2012gb, Alves:2012fx, Cheung:2012zq, Galloway:2013dma, Craig:2013fga, Chang:2014ida}.

Therefore, complete naturalness of SUSY models requires both a spectrum of light  superpartners and an extension of the MSSM Higgs sector.  This level of non-minimality has led some to argue that the price of naturalness is too high, and that Nature may prefer a simpler but more fine-tuned scenario (see \eg~\cite{Bhattacherjee:2012ed, Arvanitaki:2012ps, ArkaniHamed:2012gw}).  As discussed above, this tuning requires only a single accidental cancelation, so this point of view should be taken seriously.  However, fully exploring natural models is one of the most important tasks of particle physics.

In this spirit, the aim of this paper is to investigate the model-independence of the naturalness constraints estimated in \Eq{tuning} on the superpartner spectrum.  The naturalness bound on the stop mass can be understood on very general bottom-up
principles.  The top loop correction to the Higgs mass in the Standard Model is quadratically sensitive to UV physics.  In a natural model, some new physics must cut off this dependence.  Two possibilities are compositeness of the Higgs and/or top quark, or the existence of ``top partners,'' particles whose loop corrections cancel the quadratic UV sensitivity from top loops.  SUSY is the canonical example of the latter, with the stop playing the role of the top partner.  The bound on the stop mass in SUSY models is therefore very robust and model-independent.  The bound on the gluino mass is also quite general---it arises because $m_{\tilde{g}}$ contributes to the stop mass at loop level, which in turn contributes to the Higgs mass.\footnote{This bound can be somewhat alleviated in models where the gluino is a Dirac fermion \cite{Fox:2002bu, Kribs:2012gx}.}

There is no analogous argument for the naturalness constraint on the Higgsino mass, which is the focus of this paper.
In the MSSM and most extensions considered in the literature, a Higgsino mass $\mu$ directly contributes to the Higgs mass parameter at tree level: $\Delta m_H^2 = \mu^2$.  A Higgsino mass significantly larger than the observed Higgs mass then requires large cancelations from some other source, giving rise to fine-tuning.  This connection arises from details of the symmetry structure, so the argument for a naturalness constraint on the Higgsino mass is more model-dependent.

In this paper, we point out that the connection between Higgsino and Higgs masses is  completely severed in models where the Higgs boson is a pseudo Nambu-Goldstone boson (PNGB).  In such models, the Higgs sector has an approximate global symmetry $G$, weakly gauged by $SU(2)_W \times U(1)_Y$, that is broken spontaneously down to a subgroup $H$.  The Higgs is identified as a PNGB in the coset $G/H$ \cite{Kaplan:1983fs,Contino:2003ve,Agashe:2004rs}.\footnote{For a review and general phenomenological discussion of PNGB Higgs models, see \cite{Giudice:2007fh, Bellazzini:2014yua}.}  
Consequently, a Higgsino mass that is invariant under $G$ does not contribute to the Higgs mass, breaking the connection between the two.  As a result the Higgsino mass can naturally be much larger than $m_h$ without significant fine-tuning.

These models do however require a moderate amount of fine-tuning in the Higgs potential
to be phenomenologically viable.  Denoting the scale of $G \to H$ breaking by $f$, precision electroweak and Higgs coupling
measurements require
\beq
\frac{v^2}{f^2} \lsim 10\%,
\eeq
where $v = 246\GeV$ is the Higgs VEV.  The ratio $v^2 / f^2$ is also a direct measure of the tuning required to obtain $v \ll f$.  To make a fair comparison of the tuning in our SUSY PNGB Higgs models and conventional SUSY models, we compare this with the tuning required to raise the Higgsino mass in the MSSM, which is $\sim m_h^2 / \mu^2$.  In our models, the Higgsino mass is $\sim \la f$, where $\la$ is a dimensionless coupling in the Higgs sector that spontaneously breaks $SO(5)$.  This is also proportional to $f$, but
the tuning is reduced (compared to the MSSM with the same value of the Higgsino mass) for large $\la$.

Large couplings in the Higgs sector that spontaneously breaks $SO(5)$ are in fact required in our model to maintain the approximate $SO(5)$ symmetry while also generating a Higgs quadratic coupling sufficiently large to be compatible with the observed Higgs mass.  Moreover, they are motivated by the fact that approximate invariance under a global symmetry broken by electroweak gauge and top Yukawa couplings is natural in models with a strong coupling in the Higgs sector.  The paradigmatic example is QCD, where an approximate $SU(2)_L \times SU(2)_R$ symmetry is broken only weakly by the fact that the quarks have different electromagnetic charges and masses.

For concreteness, these ideas will be demonstrated with a simple model based on the minimal coset structure $G/H = SO(5)/SO(4)$~\cite{Agashe:2004rs}.  The only additional degrees of freedom beyond the MSSM are two gauge singlet chiral superfields that couple to the MSSM via the superpotential.  The approximate global $SO(5)$ symmetry is radiatively stable, but its origin will not be addressed in this work.  Since the model is meant as an existence proof, we will work in a subregion of the full parameter space to demonstrate that the general arguments about tuning are supported by a complete numerical analysis.

In our model, we can obtain tuning $\lsim 1/20$ for $\mu \gsim 2\TeV$.  Since we will be working with large values of $\la$, the theory has Landau poles at relatively low scales.  These can be cut off by new physics below the Landau pole scale, as is done for other SUSY models in the literature, \eg~\cite{Harnik:2003rs, Chang:2004db,Delgado:2005fq,Barbieri:2006bg}.  Alternatively, it is possible that the sector that produces the PNGB is strongly coupled, analogous to ``superconformal technicolor''  models \cite{Azatov:2011ht, Azatov:2011ps} or the ideas in~\cite{Sundrum:2009gv, Brust:2014dqa}.

A significant difference between models of the type presented below and other realistic PNGB Higgs models in the literature is that in our models the stop is the only top partner.  PNGB Higgs models without SUSY additionally require fermionic top partners in complete $SO(5)$ representations in order to render the Higgs mass calculable.  In our models, the stops are sufficient to control the quadratic corrections to the Higgs mass-squared parameter.  However, logarithmic sensitivity to UV scales remains due to the allowed $SO(5)$ breaking terms in the theory.  We check that, under renormalization group evolution, the $SO(5)$ breaking remains small up to scales in the range $10^2$--$10^6$~TeV.  This pushes the question of the origin of the $SO(5)$ symmetry beyond scales that can be presently probed.  At these high scales, the global $SO(5)$ symmetry may be the remnant of a broken gauge symmetry, for example.  The absence of fermionic top partners means that our models have fewer ingredients,
and changes the phenomenology compared to models of PNGB Higgs in the literature.

Previous studies of SUSY models with a PNGB Higgs include \cite{Redi:2010yv, Marzocca:2013fza, Parolini:2014rza,Luo:2014uha}.  In fact, the model of \cite{Marzocca:2013fza} is very similar to ours, although they include top partners to fill out complete $SO(5)$ representations.  However, to our knowledge the implications of the naturalness on the Higgsino mass have not been previously emphasized.  Also, the possibility of SUSY models with a PNGB Higgs but without $SO(5)$ top partners has not been noted.  Supersymmetric Little Higgs models have also been considered, but with the goal of eliminating tuning due to the large logarithms of the form $\log(\Lambda/m_{\rm soft})$ that appear in SUSY models \cite{Roy:2005hg,Csaki:2005fc}.  Again, these models do not focus on the naturalness implications of the Higgsino mass (for example \cite{Roy:2005hg} has $\mu \sim 200$--$400$~GeV) and incorporate fermionic top partners.  An alternative approach to SUSY breaking that does lead to heavy Higgsinos is to invoke TeV extra dimensions with Scherk-Schwarz boundary conditions \cite{Scherk:1978ta} that project out the light Higgsino \cite{Barbieri:2000vh,Dimopoulos:2014aua}.  The presence of an extra dimension means that these models require UV completion near the TeV scale, while our models are based on soft SUSY breaking.

The models considered here have potentially important implications for the interpretation of SUSY searches.  For example, suppose that at the 13 TeV LHC a SUSY signal consistent with gluino pair production followed by decays involving tops, bottoms, and a neutral LSP $\chi$ is observed, with
\begin{equation*}
m_{\tilde{g}} \sim 1.2\TeV,
\,\,
m_{\tilde{t}_2} \sim 800\GeV,
\,\,
m_{\tilde{b}_1} \sim 700\GeV,
\,\,
m_{\tilde{t}_1} \sim 600\GeV,
\,\,
m_\chi \sim 500\GeV,
\end{equation*}
where $\tilde{g}$ is the gluino, $\tilde{t}_{1,2}$ are the two stop mass eigenstates (their masses are split due to the assumption of a large $\mu$ term), and $\tilde{b}_1$ is the lighter sbottom.\footnote{There is currently a CMS search with null results for a subset of this spectrum \cite{Chatrchyan:2013iqa}.  We used {\tt Fastlim}~\cite{Papucci:2014rja} in combination with 
{\tt SUSYHIT}~\cite{Djouadi:2006bz} to check that this spectrum is plausibly allowed experimentally, and should
be readily observable at the upcoming 13~TeV run of the LHC.}  Given the large number of $\tilde{g}\,\tilde{g}$ events, approximate values for the masses and branching ratios would be inferred.  We would not know the identity of the LSP, but we know that the Higgsino mass can be no smaller than $500\GeV$, otherwise it would be the LSP.  In the MSSM, the tuning in such a spectrum would be dominated by the Higgsino contribution, and would be of order $1\%$ in the best-case scenario where the LSP is Higgsino-like.  In our model, we naturally have a much larger Higgsino mass with tuning of order
$10\%$, and the LSP would have essentially no Higgsino admixture.  Indeed, the conventional conclusion that natural SUSY is under experimental pressure relies heavily on the assumption that such spectra are tuned because of the necessity of heavy Higgsinos \cite{Evans:2013jna}.

Although we attempt to carefully quantify fine-tuning in this paper, we are not claiming that the precise value of the tuning is meaningful beyond a rough estimate.  We also do not advocate the idea that nature chooses to minimize some measure of tuning---if this were the case, SUSY would have been discovered long ago.  Our point of view is that the fact that nature is apparently somewhat tuned is a possible hint for non-minimal structure in the model, and we are exploring one such possibility.
We believe that all possible natural models
should be thoroughly examined, and it is in this spirit that we 
present our work.

The remainder of this paper is organized as follows.
In \Sref{sec:Model}, we describe our model and analyze it in several
simplified limits to elucidate the important effects that determine the
amount of tuning.
In \Sref{sec:Results}, we present the results of a complete numerical
analysis, demonstrating the improvement in tuning
relative to the MSSM.
Our conclusions are presented in \Sref{sec:Conclusions}.

\section{A SUSY PNGB Higgs Model}
\label{sec:Model}
In this section, we show that a SUSY PNGB Higgs can arise from a very simple
extension of the MSSM, namely a model with two additional gauge singlet chiral superfields.
This allows an embedding of the Higgs sector into a representation of an approximate $SO(5)$ global symmetry.  The symmetry is broken explicitly by small superpotential and soft-SUSY breaking terms, in addition to the electroweak gauge interactions and Yukawa couplings.
The model will be analyzed in several steps in order to emphasize important features:
\begin{itemize}
\item Beginning with the limit of exact $SO(5)$ symmetry, soft SUSY breaking will be introduced in order to induce a spontaneous breaking of the global symmetry, $SO(5)\rightarrow SO(4)$.  
In this limit, there are 4 massless NGBs that will be identified with the 
physical Higgs, and the Higgsinos will have mass of order the global symmetry breaking scale.
Because the Higgs potential vanishes in this limit, this already shows
that a Higgsino mass does not contribute to the light Higgs mass, 
thereby breaking the na\"ive SUSY intuition.
\item Next, tree-level terms that provide a small explicit breaking of the $SO(5)$ global symmetry are added.  Electroweak gauge symmetry is spontaneously broken, and the physical Higgs boson acquires a non-zero mass.  The model is automatically in the limit of $\tan\beta = 1$ due to an unbroken custodial $SO(4)$ symmetry.  
In this limit, the tuning required to obtain a realistic
Higgs potential can be transparently derived,
and this can be compared with the tuning for an analogous spectrum in the MSSM.
\item Next, the effect of the explicit breaking of the $SO(5)$ global symmetry by the Standard Model electroweak gauge and top Yukawa couplings will be included.  In particular, since the top Yukawa explicitly breaks $SO(4)$, it pushes the model away from the $\tan\beta = 1$ limit, generating additional 
sensitivity to the global symmetry breaking (and hence Higgsino mass) scale.  In addition, the model-dependent quadratic contributions proportional to the stop soft-mass will be included.
\item Finally, we present a complete numerical analysis of the model,
explicitly demonstrating that it exhibits the main features described above.
\end{itemize}

In order to be quantitative, we use a version of the Barbieri-Giudice measure of tuning \cite{Barbieri:1987fn}, the fractional sensitivity of the physical Higgs mass squared $m_h^2$ to changes in the various input parameters $p_i$,
\be
\Delta^{-1} =  \frac{\partial \ln m_h^2}{\partial \ln p_i}.
\ee
We have checked that the (more traditional) tuning in $v^2$ is comparable,
but we use $m_h^2$ for practical reasons.  Depending on the parameters, we find that the tuning is dominated by one of the following sources:
\begin{align}
\Delta^{-1}_\text{PNGB} \sim \frac{f^2}{v^2}; \quad\quad \Delta^{-1}_{\delta t_\beta} \sim \delta t_\beta\,\frac{\mu_{\rm eff}^2}{v^2}; \quad\quad \Delta_{\rm radiative}^{-1} \sim \frac{3}{16 \pi^2} \frac{m_{\tilde{t}}^2}{v^2} \log\left(\frac{M_{\rm SUSY}^2}{m^2_{\tilde{t}}} \right),
\end{align}
where $\delta t_\beta = \tan \beta - 1$ parameterizes the $SO(4)$ breaking,
and $\mu_\text{eff}$ is the effective $\mu$-term generated in our model.  The standard 1-loop tuning from stops is estimated by $\Delta_{\rm radiative}^{-1}$,  and sets a minimum possible tuning for a particular choice of SUSY breaking scale 
$M_\text{SUSY}$.  For sufficiently low values of $M_\text{SUSY}$ this contribution is
subdominant, and the dominant tuning is determined by a competition between the 
other two sources of tuning.

In the following, we present an approximate analytic argument for these scalings, followed by a complete numerical analysis of the model, which will demonstrate that these effects are robust.

\subsection{The $SO(5)$ Limit}
\label{sec:SO5Limit}
Our model contains the fields of the MSSM, plus two singlet chiral superfields
$\Sigma$ and $S$.  The MSSM Higgs fields $H_u$ and $H_d$ can be embedded into a fundamental representation of $SO(4)$,
\be
\Phi_i = \frac{1}{\sqrt{2}} \begin{pmatrix}
-i(H_u^1 + H_d^2) \\
H_u^1 - H_d^2 \\
i(H_u^2 - H_d^1) \\
H_u^2 + H_d^1
\end{pmatrix},
\qquad
i = 1, \ldots, 4,
\ee
where the superscripts on the $H_{u,d}$ scalars are $SU(2)_W$ indices.  This implies the relationships
\be
\Phi_i^\dagger\, \Phi_i = H_u^\dagger\, H_u + H_d^\dagger\, H_d,
\qquad
\Phi_i \,\Phi_i = - 2 \,H_u\, H_d.
\ee
Combining $\Sigma$ with these Higgs fields forms a (complex) fundamental of $SO(5)$,
\be
\Phi_a = (\Sigma, \Phi_i),
\qquad
a = 0, \ldots, 4.
\ee

We now present the model in the limit of exact $SO(5)$ symmetry.  The superpotential is
\be
W  = \frac{\lambda}{2}\, S\, \Phi_a\, \Phi_a - \frac{\kappa}{3}\, S^3.
\ee
$W$ contains no dimensionful parameters---as in the NMSSM, this can give a solution to the $\mu$ problem.\footnote{The omission of the relevant terms $S$, $S^2$, and $\Phi_a \Phi_a$, can be justified by imposing additional symmetries.}  SUSY is assumed to be broken softly by introducing the following terms into the scalar potential:\footnote{A linear term in $S$ can be forbidden by imposing additional symmetries.  $A$ terms are also neglected.}
\be
\label{eq:softpotential}
V_{\rm soft}  = m_S^2 \abs{S}^2 
+ m_\Phi^2\, \Phi_a^\dagger\, \Phi^{\vphantom\dagger}_a 
+ B_S (S^2 +\text{h.c.})
+ B_\Phi (\Phi_a\, \Phi_a + \text{h.c.}).
\ee
The couplings and soft masses can be chosen such that $S$ and $\Sigma$ acquire non-zero VEVs,
\be
\vev{S} = \frac{u}{\sqrt{2}},
\qquad
\vev{\Sigma} = \frac{f}{\sqrt{2}}.  
\ee
When $f\neq 0$, $SO(5)$ is spontaneously broken, yielding a NGB multiplet consisting of the real components of $\Phi_i$, which parameterize the coset space $SO(5)/SO(4)$.  These will be identified with the Higgs field responsible for breaking electroweak symmetry.  
The imaginary components of $\Phi_i$ make up a heavy Higgs doublet with mass
\be
\label{eq:stableminimum}
m^2_{\text{Im}(\Phi_i)} = \frac{\lambda\,\kappa\, u^2}{2} -
\frac{\lambda^2\, f^2}{4}.
\ee
This must be positive in order to have a stable vacuum.
In the absence of explicit $SO(4)$ breaking, $\tan \beta = 1$.  Furthermore, the heavy Higgs doublet has a vanishing VEV and does not mix with the light Higgs.

In addition to the Higgs doublets, the spectrum also contains two singlet scalars and two singlet pseudoscalars, which are admixtures of the real and imaginary components of $S$ and $\Sigma$.  
In the limit of vanishing $B$-terms
$B_S, B_\Phi \to 0$ the theory above has a $U(1)_R$ symmetry with
charges $R(S) = R(\Phi) = \frac 23$, and therefore an associated axion-like NGB.
Nonzero $B$-terms are therefore important for the phenomenology of the theory,
but do not significantly affect the aspects of the Higgs sector that are the focus of this paper.  For the numerical analyses performed below, parameters will be chosen such that this axion state is lifted.

The Higgsino mass is given by
\be
\mu_{\rm eff} = \frac{\lambda\, u}{\sqrt{2}},
\ee
while the NGBs are exactly massless.  Therefore, this simple limit already demonstrates the separation between the (so-far massless) Higgs scalars and the Higgsinos.  

\subsection{Explicit Breaking of $SO(5)$}
\label{sec:ExplicitSO5Breaking}
We now include terms that explicitly break the global $SO(5)$ symmetry at tree-level in order to generate a Higgs potential.
The next subsection will analyze the largest effects from the loops of Standard Model particles/sparticles.
Our purpose is to discuss the tuning required for realistic electroweak symmetry
breaking in the simplest possible context.

Considering only dimensionless terms in the superpotential, we include the following $SO(5)$-breaking couplings
\be
\label{eq:DeltaW}
\Delta W = \frac{\lambda'}{2}\, S \, \Sigma^2
+ \frac{\eta}{2} \,S^2\, \Sigma
- \frac{\kappa'}{2}\, \Sigma^3,
\ee
and the soft SUSY- and $SO(5)$-breaking terms
\be
\label{eq:DeltaV}
\Delta V_{\rm soft} = \Delta m_\Sigma^2\, |\Sigma|^2 
+ \Delta m_{S \Sigma}^2 (S\, \Sigma^\dagger + \text{h.c.})
+ B_\Sigma\, (\Sigma^2 + \text{h.c.}) 
+ B_{S \Sigma} \, (S\, \Sigma + \text{h.c.})
\ee
in the potential.
In order to understand the light Higgs potential, it is simplest to use the effective theory below the mass scale of the heavy fields.  Due to the large separation of scales, it is consistent to work at leading order in the $SO(5)$ breaking terms above.
We analyze this model in the limit where the $B$-terms and $\Delta m_{S \Sigma}^2$ vanish, purely for simplicity. 
However, we include $B_S$ in our numerical analysis below, and also discuss the radiative stability of neglecting various terms in the appendix.
Explicit $SO(5)$ breaking gives rise to a potential for the light Higgs doublet
$H$ with the Standard Model form
\be
V = m_H^2\, H^\dagger\, H + \frac{\lambda_H}{4} (H^\dagger\, H)^2.
\ee
The effective parameters $m_H^2$ and $\lambda_H$ can be obtained by
integrating out the heavy fields at tree level:
\be
m_H^2 = -\Delta m_\Sigma^2 - \frac{\lambda'}{4} \left[ \lambda \, f^2
- 2(\kappa - 2\,\lambda) u^2 \right]
- \frac{\eta}{4} \left[ \lambda\, f\, u - (2\,\kappa - \lambda) \frac{u^3}{f} \right]
+ \frac{3\, \kappa'}{2} \lambda\, f\, u,
\ee
and
\be
\lambda_H = \frac{\eta}{2} \left[ \frac{(2\,\kappa - \lambda) u^3}{f^3} -
\frac{\lambda \, u}{2\, f} \right] - \frac{3\, \kappa' \, \lambda \, u}{f}.
\ee
Here we have traded the $SO(5)$ invariant parameters $m_S^2$ and $m_\Phi^2$
for the VEVs $f$ and $u$.
Note that the terms in the light Higgs potential are proportional to explicit $SO(5)$ breaking,
as they must be.
There is clearly sufficient freedom to obtain $m_H^2 < 0$ and $\lambda_H > 0$, while still having a stable minimum (see \Eref{eq:stableminimum}).

To understand the tuning, consider the simplest case where $\lambda \sim \kappa$,
$u \sim f$, and $\eta \sim \lambda' \sim \kappa' \ll \lambda$.
Then
\be
m_H^2 \sim -\Delta m_\Sigma^2 + \eta\, \lambda \, f^2,
\qquad
\lambda_H \sim \eta \, \lambda.
\ee
The VEV of the Higgs field is given by
\beq
v^2 = -\frac{m_H^2}{\lambda_H},
\eeq
such that $v \gsim f$ unless $m_H^2$ is tuned to be smaller than its natural
size.
This is the canonical tuning inherent in PNGB Higgs models.  For example, if the small value of $m_H^2$ is obtained by canceling 
$\Delta m_\Sigma^2$ against the other terms, the largest sensitivity is given by
\beq
\Delta^{-1}_\text{PNGB}
\simeq -\frac{\Delta m_\Sigma^2}{m_H^2}
\sim \frac{f^2}{v^2}.
\eeq
The tuning is always of order $v^2 / f^2$, which can be understood from the fact
that we are tuning the Higgs to be light compared to heavy states whose mass
is proportional to $f$.
The same parameter $v^2 / f^2$ controls the deviation of the couplings of the
light Higgs from the SM values.
The observed Higgs couplings imply $v^2 / f^2 \lsim 10\%$, so there is an unavoidable
tuning of order $10\%$ in this framework.

It may appear that this tuning of the Higgs potential spoils our claim of enhanced
naturalness for this model.
In fact, the Higgsino mass is also proportional to $f$, so the tuning required to raise the Higgsino mass in the MSSM is proportional to $v^2/f^2$, just like the tuning in the Higgs potential above.
However, the tuning in our model is parametrically improved relative to an MSSM-like theory when the 
dimensionless $SO(5)$ preserving couplings are large.
In an MSSM-like theory where the Higgsino mass contributes to the Higgs mass at
tree level, we have a tuning
\be
\Delta^{-1}_{\rm MSSM} \simeq \abs{\frac{\mu_{\rm eff}^2}{m_H^2}}
= \frac{2\, \mu_{\rm eff}^2}{m_h^2},
\ee
where $m_h = 125$~GeV is the physical Higgs mass.
Note that this is a conservative estimate of the tuning, since it assumes that
there is a natural mechanism for generating the Higgs quartic and does not include the tuning contribution from stop loops.
For example, this is the case in the NMSSM for large values of the $S\, H_u\, H_d$ 
coupling (\emph{i.e.}~``$\lambda$-SUSY'' \cite{Barbieri:2006bg}),
but not for the MSSM where stop loops generate the Higgs quartic.
To make a fair comparison to our model, we consider
the ratio of the tuning in our model to the tuning in an MSSM-like theory 
for the same value of the Higgsino mass, $\mu_{\rm eff} \sim \la u$, yielding
\be
\label{eq:Tscaling}
T \equiv \frac{\Delta_{\rm PNGB}^{-1}}{\Delta_{\rm MSSM}^{-1}}
\sim \frac{\la^2}{\la_H},
\ee
where $\la$ is a $SO(5)$-preserving coupling in our model
and $\la_H = m_h^2 / 2 v^2 = 0.13$ is the SM Higgs coupling.
We see that the tuning is parametrically improved relative to an MSSM-like
model.  This is one of the main conclusions of our work.

The tuning in theories with heavy Higgsinos is
most significantly improved relative to MSSM-like theories when the Higgs sector that spontaneously breaks $SO(5)$ is strongly coupled.
This means that we get the maximum gain in a limit where our model is not
calculable.
Our attitude toward this is that the model we are presenting is an existence
proof, and we will use it to demonstrate that the above simple picture of the
tuning can be realized in a complete model.
The improvement in the tuning in this model is then limited by the
weak coupling that we need for calculability, but we can be confident that
there are no hidden problems in this scenario.
It is therefore plausible that there are strongly-coupled models with the
same general features, and we will illustrate the possibilities by extrapolating
the present model all the way to strong coupling.

\subsection{Explicit $SO(5)$ Breaking by Gauge and Yukawa Couplings
\label{sec:ExplicitSMBreaking}}

In this section, the effects of the Standard Model electroweak gauge and Yukawa couplings is discussed.
These must not be too large if we are to maintain a light Higgs mass without fine tuning.  Not surprisingly, we find that the largest contribution comes from top/stop loops. 

First, consider the electroweak gauge couplings.  Because $\tan \beta \simeq 1$, the electroweak $D$-term 
contribution to the  Higgs potential is negligible.  However, electroweak loops will generate a radiative correction to the mass-squared parameters for $H_u, H_d$ but not $\Sigma$,
thereby breaking $SO(5)$.  The largest contribution comes from the Higgsinos due to the large Higgsino
mass in our models.  The value is model-dependent because  
it is sensitive to UV-scale physics, but it can be estimated using the leading-log approximation:
\be
\label{eq:radiativeEWcontribution}
(\Delta m_{\Sigma}^2)_{\rm EW} \simeq - \frac{3\, g_2^2}{8\, \pi^2} \,\mu_{\rm eff}^2 \log \left(\frac{M_{\rm SUSY}}{\mu_{\rm eff}}\right)
\ee
where $M_{\rm SUSY}$ is the SUSY-breaking scale and $g_2$ is the $SU(2)_W$ gauge coupling.  
Therefore, requiring small $\Delta m_{\Sigma}^2$ and the approximate $SO(5)$ symmetry potentially gives rise to a loop-level tuning.  
Using \Eref{eq:radiativeEWcontribution}, one expects
\be
\label{eq:EWradiativetuning}
\Delta_{\rm radiative}^{-1} \sim \frac{3\, g^2}{16\, \pi^2} \frac{\mu_{\rm eff}^2}{v^2} \log\left(\frac{M_{\rm SUSY}}{\mu_{\rm eff}}\right).
\ee
In our numerical analysis below, we include this tuning by including in the potential
\be
\Delta m_\Sigma^2 = \Delta m_{\Sigma, 0}^2 + \left(\Delta m_\Sigma^2\right)_{\rm EW}
\ee
and calculate the tuning with respect to $\Delta m_{\Sigma, 0}^2$.  Therefore, any radiative tuning of \Eref{eq:EWradiativetuning} will show up as the tuning of $\Delta m_{\Sigma, 0}^2$ against the Higgsino loop contribution that is required to keep $\Delta m_\Sigma^2$ small.  We find that this tuning is generally subdominant in the regions of parameter space we consider, but it can become relevant for larger Higgsino masses.  

Another important source of explicit $SO(5)$ breaking is the large top Yukawa coupling, which is particularly important as it impacts the potentials for $H_u$ and $H_d$ differently. Integrating out the tops and stops generates a correction to the Higgs potential for $H_u$ but not for $H_d$
\be
\Delta V_t = \Delta m_{H_u}^2 H_u^\dagger \,H_u 
+ \frac{\Delta \lambda_{H_u}}{4} \big(H_u^\dagger\, H_u\big)^2.
\ee
These contributions break the custodial $SO(4)$ subgroup of the approximate global $SO(5)$, so this source of explicit breaking can be parameterized in terms of the resulting deviation from $\tan \beta = 1$.  In particular, for $\delta t_\beta = \tan \beta - 1 \neq 0$, the light Higgs mass will pick up contributions $\Delta m_H^2 \sim \delta t_\beta \,\mu_{\rm eff}^2$.  This can be understood as a mixing between the PNGB Higgs and the second Higgs doublet, with the amount of mixing set by $\delta t_\beta$, and this effect induces an additional source of sensitivity to the global symmetry-breaking scales $u, f$.  As a result, the corresponding tuning is
\be
\Delta^{-1}_{\delta t_\beta} \sim \delta t_\beta\,\frac{\mu_{\rm eff}^2}{v^2} \propto \Delta^{-1}_{\rm MSSM}
\ee
such that $T \propto \delta t_\beta$ is approximately constant  for regions of the PNGB sector parameter space where this contribution to the tuning dominates.  Hence, for a given value of $\delta t_\beta$, there will be a maximum possible improvement in the tuning with respect to the MSSM.

The contribution from top/stop loops to the Higgs quartic is
\cite{Martin:1997ns}
\begin{align}
\Delta \lambda_{H_u} = \frac{3}{\pi^2} \left(\frac{m_t}{v_u}\right)^4 & \left\{\log\left(\frac{m_{\tilde{t}_1} \,m_{\tilde{t}_2}}{m_t^2}\right) + c_{\tilde{t}}^2\, s_{\tilde{t}}^2\, \frac{m_{\tilde{t}_2}^2-m_{\tilde{t}_1}^2}{m_t^2} \log\left(\frac{m_{\tilde{t}_2}^2}{m_{\tilde{t}_1}^2}\right) \right. 
\nonumber\\ 
& \qquad \left. + c_{\tilde{t}}^4\, s_{\tilde{t}}^4\, \frac{(m_{\tilde{t}_2}^2 - m_{\tilde{t}_1}^2)^2 - \frac{1}{2} (m_{\tilde{t}_2}^4 - m_{\tilde{t}_1}^4) \log(m_{\tilde{t}_2}^2/m_{\tilde{t}_1}^2)}{m_t^4}\right\}.
\label{eq:lambdaHuStopLoops}
\end{align}
In a natural model with lighter stops, as envisioned here, this contribution to $\lambda_H$ is subdominant for $m_h = 125 \GeV$.  Nonetheless, it does yield an $\sim 30\%$ contribution to the Higgs mass, so we include a contribution to $\Delta \lambda_{H_u}$ given by \Eref{eq:lambdaHuStopLoops} in our numerical analysis below.

As for the electroweak correction discussed above, the term $\Delta m_{H_u}^2$ is model-dependent since it is sensitive to UV-scale physics.  
In the leading-log approximation, the size of the radiative contribution to 
$\Delta m_{H_u}^2$ is given by
\be
\label{eq:mHuStopLoops}
\left(\Delta m_{H_u}^2\right)_{\rm stop} \simeq
-\frac{3\, y_t^2}{16 \, \pi^2} \Big(m_{Q_3}^2 + m_{u_3}^2\Big)
\log\left(\frac{
M_{\rm SUSY}^2
}{m_{\tilde{t}_1} m_{\tilde{t}_2}}\right),
\ee
where we have neglected $A$ terms and the contribution from the gluino for simplicity.  Thus, a reasonable concern is that maintaining small $\delta t_\beta$ (and hence improvement relative to the MSSM) represents an additional source of tuning in the model due to the radiative (in)stability of small $\Delta m_{H_u}^2$.  The magnitude of the radiative tuning can be estimated using \Eref{eq:mHuStopLoops},
\be
\label{eq:topradiativetuning}
\Delta_{\rm radiative}^{-1} \sim \frac{3\, y_t^2}{32\, \pi^2} \frac{m_{Q_3}^2}{v^2} \log\left(\frac{M_{\rm SUSY}^2}{m_{\tilde{t}_1} m_{\tilde{t}_2}}\right).
\ee
This contribution gives a lower bound on the tuning in the model for a given set of stop masses and 
$M_\text{SUSY}$, as in any natural SUSY model.
Note that we do not include $\Delta^{-1}_{\rm radiative}$ as a contribution to $\Delta^{-1}_\text{MSSM}$ when computing $T$, making our comparison to the MSSM maximally conservative.
In our numerical analysis below, we account for this tuning by including in the potential
\be
\Delta m_{H_u}^2 = \Delta m_{H_u, 0}^2 + \left(\Delta m_{H_u}^2\right)_{\rm stop}
\ee
and calculate $\Delta^{-1}$ with respect to both $\Delta m_{H_u, 0}^2$ and the stop mass-squared parameters.    

The stop mass contributions described above become more important for
large $\mu_\text{eff}$ because this increases the stop mass splittings.  This increases the tension with naturalness because
larger soft-masses in the stop sector are required to avoid conflict with LHC bounds on the lightest stop.  One could allow $A_t \ne 0$, but this would also contribute to $\left(\Delta m_{H_u}^2\right)_{\rm stop}$ through the renormalization group evolution.  In the numerical results below, we will fix the lowest stop mass to be 600 GeV, thereby accounting for the increased contribution to the tuning with larger soft-masses.

As one might expect, we find that preserving the gain relative to the MSSM requires maintaining the limit where the PNGB description is approximately valid, which includes the requirement that $\tan\beta \simeq 1$.  Consequently, while the largest gains relative to the MSSM are achieved for strong coupling, the improvement does not increase arbitrarily as $\lambda \rightarrow \infty$.  As $\lambda \propto \mu_{\rm eff}$ increases, small $\delta t_\beta$ is required such that $\Delta_{\delta t_\beta}^{-1}$ does not dominate the tuning.  Eventually, however, the tuning $\Delta_{\rm radiative}^{-1}$ required to keep $\Delta m_{H_u}^2$ (and hence $\delta t_\beta$) small prevents $\delta t_\beta$ from being decreased further.  In other words, the top/stop loops are what limit the improvement in tuning for large $\la$.

\section{Results}
\label{sec:Results}
In this section, we confirm the above results based on a full numeric analysis of the tree-level 
potential, including the most important radiative corrections \Erefs{eq:EWradiativetuning}{eq:topradiativetuning} as described above.  A weakly-coupled benchmark point is presented as a proof of principle, and we also present results as a function of $\la$ for a subspace of the parameter space in order to demonstrate the improvement when extrapolating to strong coupling.

The most important experimental constraint on this model comes from Higgs coupling measurements
at the LHC.
In our model, the light Higgs mixes with CP-even components of $S$ and $\Sigma$,
resulting in a universal reduction factor $\kappa_h$ 
in the couplings between the couplings of the
Higgs to all other SM particles.
Using the full 7 and 8 TeV data sets, ATLAS has given the constraint \cite{ATLASHiggsConstraints}
\beq
\label{eq:kappahConstraint}
\kappa_h - 1 > -0.064.
\eeq
As in any PNGB Higgs model, we have $\kappa_h - 1 \sim v^2 / f^2$,
assuming $v \ll f \sim u$.
In the minimal composite Higgs model the constraint \Eq{eq:kappahConstraint}
implies $f > 710\GeV$, which in turn requires tuning of order $v^2 / f^2 \sim 10\%$.
In our model, the precise constraint has a more complicated parametric dependence because the coupling suppression arises from the mixing of 3 states,
but the conclusions are essentially the same.
We find that imposing the constraint \Eq{eq:kappahConstraint} implies tuning
of at least $10\%$.

\subsection{Benchmark Model}
We present a complete weakly coupled benchmark in \Tref{tab:bench} that satisfies all experimental constraints.
This will be used as a starting point to extrapolate into the strongly-coupled regime in the following subsection.
Note that we consider only a subset of the possible couplings for simplicity.  In the appendix we show that this choice is sufficiently radiatively stable that it does not introduce additional tuning.

\begin{table}[h!]
\footnotesize
{\renewcommand{\arraystretch}{1.5}
\centering
\setlength{\tabcolsep}{1.5em}
\begin{tabular}{|c|c|c|c|c|c|}
\hline
\multicolumn{6}{|c|}{$SO(4)$ symmetric input parameters} \\
\hline
$\lambda$ & $\kappa$ & $u$ [TeV] & $f$ [TeV] & $B_S$  $\big[\text{TeV}^2\big]$ & $\eta$ \\
\hline
$1.5$ & $2.0$ & $1.1$ & $0.65$ & $-0.04$ & $0.0872$\\
\hline
\end{tabular} \\
\vspace{0.25cm}
\begin{tabular}{|c|c|c|c|}
\hline
\multicolumn{4}{|c|}{Stop sector input parameters} \\
\hline
$\tan \beta$ & $m_{Q_3} = m_{u_3}$ [TeV] & $m_{d_3}$ [TeV] & $A_t = A_b$ [TeV] \\
\hline
$1.05$ & $0.718$ & $1.5$ & 0 \\
\hline
\end{tabular} \\
\vspace{0.25cm}
\begin{tabular}{|c|c|c|c|}
\hline
\multicolumn{4}{|c|}{Soft SUSY-breaking masses} \\
\hline
$m_S^2$ $\big[\text{TeV}^2\big]$ & $m_\Phi^2$  $\big[\text{TeV}^2\big]$ & $\Delta m_{\Sigma, 0}^2$  $\big[\text{TeV}^2\big]$ & $\Delta m_{H_u, 0}^2$  $\big[\text{TeV}^2\big]$ \\ \hline
$-4.48$ & $0.216$ & $0.103$ & $0.174$ \\
\hline
\end{tabular}}
\caption{Input parameters for a weakly-coupled benchmark.  The upper table lists the $SO(4)$ preserving parameters as discussed in Sections~\ref{sec:SO5Limit} and~\ref{sec:ExplicitSO5Breaking}, and the middle table lists the parameters of the stop sector,
which violate $SO(4)$ and are discussed in Section~\ref{sec:ExplicitSMBreaking}.  The lower table shows the soft SUSY-breaking masses that give rise to the
assumed VEVs.}
\label{tab:bench}
\end{table}

The physical masses of the particles are
\begin{align}
m_{\tilde{t}} & = 600 \GeV,\ 851 \GeV, \\
m_{\tilde{b}} & = 718 \GeV,\ 1.5 \TeV, \\
m_h & = 125 \GeV,\\
m_{\tilde{H},\,H^0,\,H^\pm,\,A^0} &= 1.73 \TeV, \\
\text{scalars: } m & = 811 \GeV,\ 3.1 \TeV, \\
\text{pseudoscalars: } m & = 337 \GeV,\ 2.29 \TeV.
\end{align}

The tunings with respect to the input parameters are as follows:
{\renewcommand{\arraystretch}{1.5} \setlength{\tabcolsep}{0.8em} \begin{center}
\begin{tabular}{|c||c|c|c|c|c|c|c|c|c|} \hline
$p_i$ & $\lambda$ & $\kappa$ & $m_S^2$ & $m_\Phi^2$ & $\Delta m_{\Sigma, 0}^2$ & $\eta$ & $\Delta m_{H_u, 0}^2$ & $m_{Q_3}^2$ & $m_{u_3}^2$ \\ \hline
$\Delta^{-1}$ & 22 & 29 & 4 & 4 & 12 & 8 & 10 & 9 & 9 \\ \hline
\end{tabular}
\end{center}}
\noindent The tunings in $\lambda, \kappa$ capture $\Delta^{-1}_{\rm PNGB}$, 
while the tunings in $\Delta m_{\Sigma, 0}^2, \Delta m_{H_u, 0}^2, m_{Q_3}^2$ and $m_{u_3}^2$ 
correspond to the radiative tunings described in \Erefs{eq:EWradiativetuning}{eq:topradiativetuning}.  Note that, for $M_{\rm SUSY} = 10^4~\big(10^6\big) \TeV$, we would have
$\Delta m_{H_u, 0}^2 = 0.479~(0.785) \TeV^2$ and 
$\Delta^{-1}_{\rm radiative} \sim 25~(50)$. 
As expected, the radiative tuning required to keep $\tan \beta \simeq 1$ dominates for larger values of $M_{\rm SUSY}$.

This benchmark model is tuned at the level of $3\%$.  
Given that the Higgsino mass is $\mu_{\rm eff} = 1.17 \TeV$, the corresponding Higgsino tuning in an 
MSSM-like model would be $\sim 0.5 \%$, corresponding to an improvement factor $T = 5.9$ (see \Eq{eq:Tscaling}).
This modest improvement is expected since we are working at weak coupling.  

\subsection{Strong Coupling}
We now extrapolate the results of the benchmark model to larger values of $\lambda$.
We do this by simply using the same approximations made for the weakly-coupled model.
Even though some quantities will have corrections of order $100\%$, we expect that the qualitative
estimates of the tuning are accurate.
The other input parameters are fixed as follows: $f = 650 \GeV$, $u = 1.1 \TeV$, 
$\kappa = \frac 43 \lambda$, $B_S = -0.04 \TeV^2$.
The parameters $\Delta m_{\Sigma, 0}^2$ and $\eta$ are chosen to reproduce 
$v = 246\GeV$ and $m_h = 125\GeV$.

\begin{figure}[h!]
\centering
\vspace*{20pt}
\includegraphics[width=.45 \textwidth]{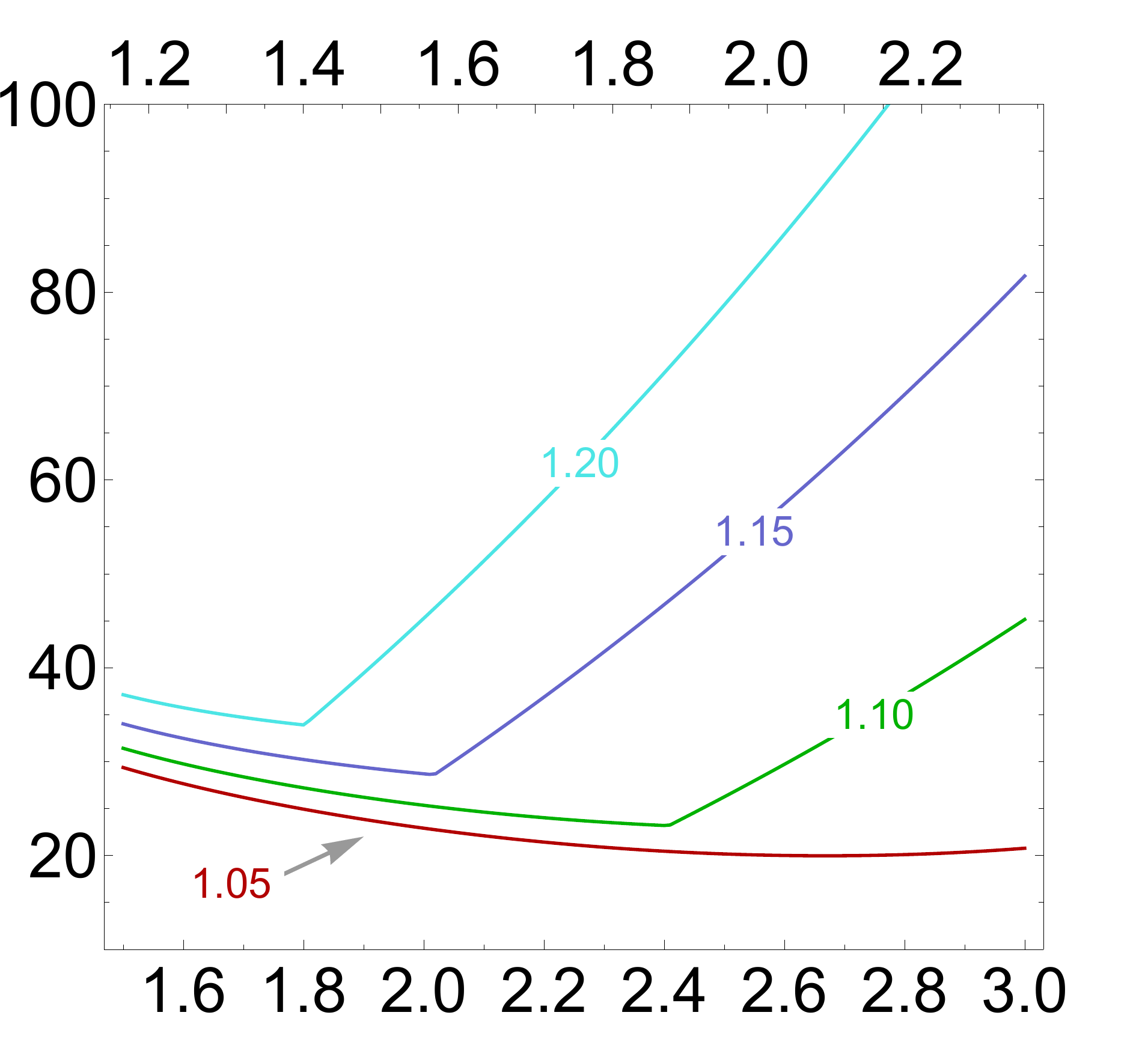} 
\hspace{-0.02\textwidth}
\includegraphics[width=.45 \textwidth]{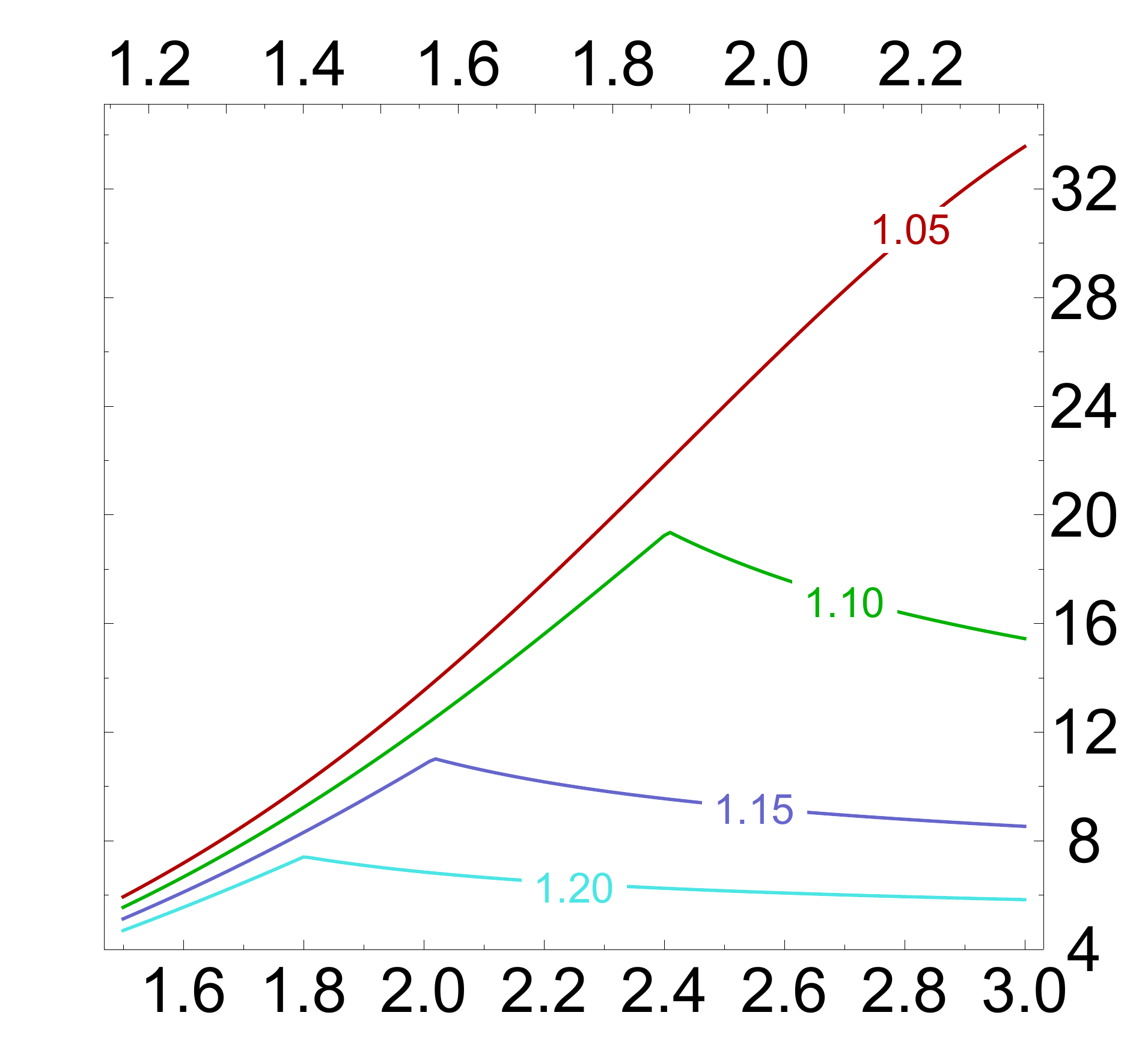} 
\hspace*{20mm}
\begin{minipage}{0cm}
\vspace*{-15.1cm}\hspace*{-8.8cm}{\large $\mu_{\rm eff} \,\,[\text{TeV}]$}
\end{minipage}
\begin{minipage}{0cm}
\vspace*{.1cm}\hspace*{-8.8cm}{\large $\lambda$}
\end{minipage}
\begin{minipage}{0cm}
\vspace*{-8.cm}\hspace*{-17.7cm}\rotatebox{90}{{\large $\Delta_{\rm max}^{-1}$}}
\end{minipage}
\begin{minipage}{0cm}
\vspace*{-15.1cm}\hspace*{1.7cm}{\large $\mu_{\rm eff} \,\,[\text{TeV}]$}
\end{minipage}
\begin{minipage}{0cm}
\vspace*{.1cm}\hspace*{2.4cm}{\large $\lambda$}
\end{minipage}
\begin{minipage}{0cm}
\vspace*{-7.6cm}\hspace*{6.3cm}\rotatebox{270}{{\large $T$}}
\end{minipage}
\vspace*{5pt}
\caption{The left (right) panel shows $\Delta_{\rm max}^{-1}$ ($T$) as a function of 
$\lambda$ for different values of $\tan \beta$ given on each curve.  
The other fixed parameters are given in the text.}
\label{fig:T}
\end{figure}

\begin{figure}[h!]
\centering
\vspace*{20pt}
\includegraphics[width=.45 \textwidth]{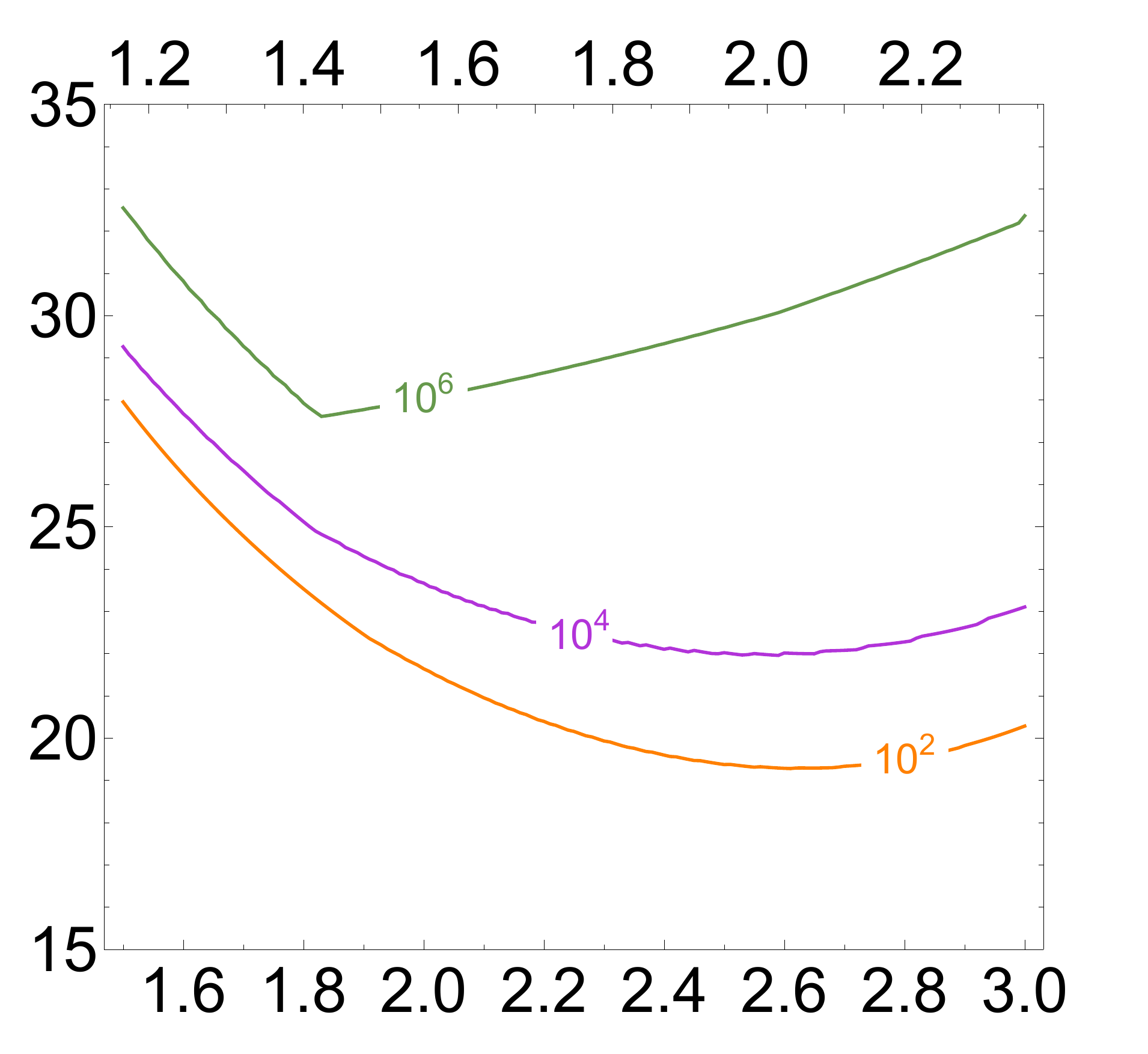} 
\hspace{-0.02\textwidth}
\includegraphics[width=.45 \textwidth]{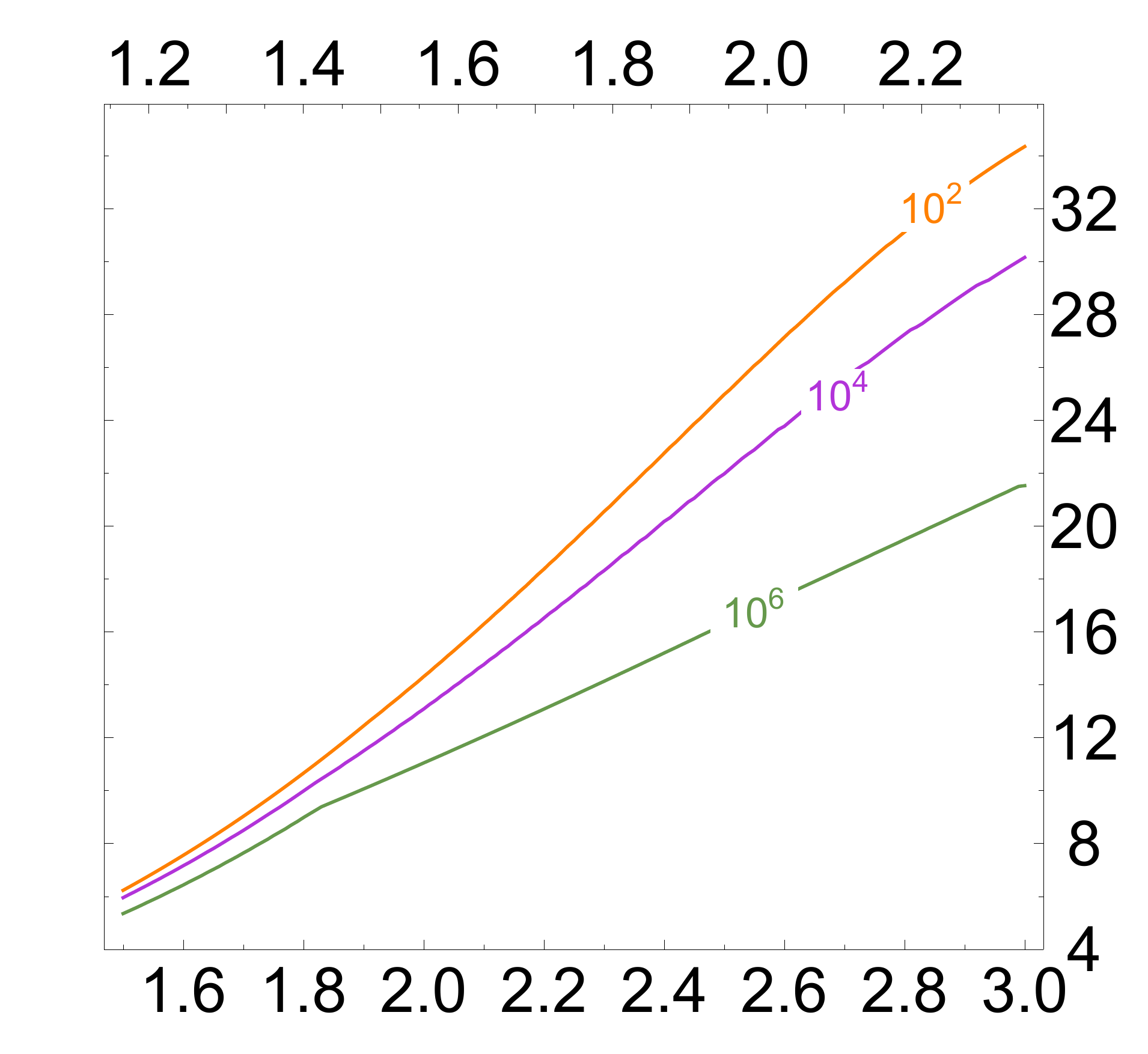} 
\hspace*{20mm}
\begin{minipage}{0cm}
\vspace*{-15.1cm}\hspace*{-8.8cm}{\large $\mu_{\rm eff} \,\,[\text{TeV}]$}
\end{minipage}
\begin{minipage}{0cm}
\vspace*{.1cm}\hspace*{-8.8cm}{\large $\lambda$}
\end{minipage}
\begin{minipage}{0cm}
\vspace*{-8.cm}\hspace*{-17.7cm}\rotatebox{90}{{\large $\Delta_{\rm max}^{-1}$}}
\end{minipage}
\begin{minipage}{0cm}
\vspace*{-15.1cm}\hspace*{1.7cm}{\large $\mu_{\rm eff} \,\,[\text{TeV}]$}
\end{minipage}
\begin{minipage}{0cm}
\vspace*{.1cm}\hspace*{2.4cm}{\large $\lambda$}
\end{minipage}
\begin{minipage}{0cm}
\vspace*{-7.6cm}\hspace*{6.3cm}\rotatebox{270}{{\large $T$}}
\end{minipage}
\vspace*{5pt}
\caption{The left (right) panel shows $\Delta_{\rm max}^{-1}$ ($T$) as a function of $\lambda$, 
for different values of $M_\text{SUSY}$ given on each curve.  
At each point, $\tan \beta$ has been chosen to minimize the tuning.
The other fixed parameters are given in the text.}
\label{fig:marginalized}
\end{figure}

Our results are presented in Figs.~\ref{fig:T} and \ref{fig:marginalized}.  Both figures show the largest tuning in our model in the left panel, and the ratio of tuning compared to an MSSM-like model in the right panel.  The purpose of \Fref{fig:T} is to show how the simple scaling arguments provided above are reproduced in the full numerical analysis.  In particular, four different choices of $\tan \beta$ are shown.  The transition from a regime where the tuning is dominated by $\Delta^{-1}_\text{PNGB}$ to where it is dominated by $\Delta^{-1}_{\delta t_\beta}$ is manifest from the turn over.  Note that, for the curve with $\tan \beta = 1.05$, $\Delta^{-1}_{\delta t_\beta}$ does not dominate for the values of $\lambda \propto \mu_{\rm eff}$ shown.
Here, we have fixed $M_{\rm SUSY} = 10^2 \TeV$; the radiative tuning is always subdominant for this choice of parameters.

\Fref{fig:marginalized} presents the bottom line results for the tuning in the
model at stronger coupling.
It shows $\Delta^{-1}_{\rm max}$ and $T$ for three choices of $M_{\rm SUSY}$, with 
$\tan \beta$ chosen at each point to minimize the tuning.
The quadratic improvement of the tuning as $\lambda$ increases turns over at large 
values of $\lambda$ due to the radiative tuning, which is more important at larger
values of $M_\text{SUSY}$.  This shows that a significant improvement in tuning with respect to MSSM-like models is possible in this framework.


\section{Conclusions}
\label{sec:Conclusions}
We have presented a general mechanism for improved naturalness in SUSY theories with heavy Higgsinos.
The main idea is that if the Higgs is a pseudo-Nambu-Goldstone boson (PNGB) arising from a global symmetry breaking pattern such as $SO(5) \to SO(4)$, an $SO(5)$ preserving Higgsino mass does not contribute to the Higgs mass, decoupling the Higgsino mass from the naturalness of the Higgs mass at tree level.
This implies that experimentally allowed models with relatively light stops and LSP masses not far below the stop masses can be natural.

We presented a simple model that realizes this scenario.
The only fields beyond the MSSM that are required are two additional singlet chiral
multiplets.  In particular, there are no additional top partners to fill out $SO(5)$ multiplets.
The model has an approximate $SO(5)$ global symmetry, and the observed Higgs is a 
PNGB associated with the spontaneous breaking of this symmetry.
A natural and phenomenologically-acceptable Higgs potential can arise from a
combination of top/stop loops and tree-level $SO(5)$ breaking.

The tuning in this model was compared with an MSSM-like model where the Higgsino mass
contributes to the Higgs mass at tree level.
We find a parametric enhancement of naturalness when the couplings in the Higgs sector that spontaneously breaks $SO(5)$ are large.
A model with a Higgsino mass of $2 \TeV$ would lead to tuning $\sim 1/20$ in this model,
an improvement with respect to the Higgsino tuning in an MSSM-like model by a factor 
of up to $30$.
Our model explains the observed Higgs mass of $125\GeV$ without additional tuning,
while the MSSM-like model would require either additional structure
(such as the NMSSM) or additional tuning to accomplish this.

Only experiment can tell us whether the electroweak scale is fine-tuned.  
In the meantime, we must continue to explore all possible angles of electroweak naturalness.  The aim
of this paper is to remind us that the only truly model independent naturalness constraints on 
the masses of superpartners come from radiative corrections to the Higgs mass.
In particular, the Higgsino mass can be naturally large in 
SUSY theories minimally extended beyond the MSSM.

\pagebreak
\section*{Acknowledgements}
We thank David Pinner and Josh Ruderman for useful conversations.  
TC is supported by DoE contract number DE-AC02-76SF00515 and by an LHC Theory Initiative 
Postdoctoral Fellowship, under the National Science Foundation grant PHY-0969510. 
TC thanks the KITP in Santa Barbara where some of this research was performed,
and for the support from the National Science Foundation under Grant No. NSF PHY11-25915. 
TC also thanks the MITP in Mainz where additional work was performed. 
JK is supported by the DoE under contract number DE-SC0007859 and Fermilab, operated by Fermi Research Alliance, LLC under contract number DE-AC02-07CH11359 with the United States Department of Energy.  
TC and JK thank the CFHEP in Beijing where some of this research was performed.  
MAL is supported by the Department of Energy under grant DE-FG02-91ER406746.

\appendix
\section*{Appendix: Additional Symmetry-Breaking Terms}
\label{sec:AdditonalTerms}

In this appendix, we elaborate on the simplifying assumptions made in \Sref{sec:Results}.  We considered a minimal set of parameters including soft SUSY-breaking masses that preserve both $SO(5)$ and $U(1)_R$ ($m_S^2, m_\Phi^2$), that explicitly break $SO(5)$ ($\Delta m_\Sigma^2, \Delta m_{H_u}^2$) and that explicitly break $U(1)_R$ ($B_S$), in addition to a single $SO(5)$-violating superpotential coupling ($\eta$).  These parameters are sufficient to yield a viable model, and so we focused on them for simplicity, although other terms could have been included.  Indeed, some of these terms will be generated radiatively, and so must be included in a complete analysis.

We have confirmed numerically that including additional terms does not disrupt the stability of our solutions.  Moreover, due to non-renormalization of the superpotential and the fact that any radiative corrections must be proportional to small symmetry-breaking parameters, the neglected terms can be consistently treated as small perturbations to the above setup without introducing sizable radiative tuning.  This allows us to consistently neglect the couplings $\lambda^\prime, \kappa^\prime$ in \Eref{eq:DeltaW}.  Similarly, consider (for example) the soft SUSY-breaking mass $m_{S \Sigma}^2$ in \Eref{eq:DeltaV}.
This term breaks $SO(5)$, and as such will receive radiative contributions proportional to $\eta$ and large soft SUSY-breaking masses.  For instance, $S$ loops will generate corrections of size
\be
\label{eq:deltamSSigma}
\Delta m_{S \Sigma}^2 \sim \frac{\lambda\, \eta}{16 \,\pi^2} \, m_S^2.
\ee
Such a contribution is a loop factor smaller than the leading contributions to $m_H^2 \sim \lambda \, \eta \, f^2 \sim \lambda \, \eta \,m_S^2$, allowing $m_{S \Sigma}^2$ to be taken small enough such that its contribution to the Higgs potential is subdominant without simultaneously introducing large radiative tuning.

In addition, we considered a non-zero $B$-term, $B_S$, in order to lift the $U(1)_R$ flat direction and to give mass to the corresponding NGB---this could also have been accomplished with a non-zero $B_\Phi$ term, see \Eref{eq:softpotential}.  As both terms are $SO(5)$-preserving, they do not significantly influence the details of the Higgs potential.  Furthermore, as $U(1)_R$ is only softly broken by these terms, radiative corrections must be proportional to $B_{S, \Phi}$, such that the smallness of these terms relative to the other large mass scales in the model (namely $m_S^2$ and $m_\Phi^2$) is radiatively stable.  While $B_{\Sigma, S \Sigma}$ in \Eref{eq:DeltaV} would influence the Higgs sector, such terms require explicit breaking of both $SO(5)$ and $U(1)_R$, so can be kept small without introducing significant loop-level tuning.
Finally, we have neglected the possible inclusion of $A$ terms.  Again, small $A$ terms are radiatively stable as such terms must be proportional to soft SUSY-breaking and $U(1)_R$-breaking, and be linear in mass---as we have no such terms, large $A$ terms will not be generated.


\end{spacing}
\begin{spacing}{1.1}
\bibliography{NaturalSUSYWithoutHiggsinos}
\bibliographystyle{utphys}
\end{spacing}
\end{document}